\documentclass[a4paper,12pt]{article}

\usepackage{amsmath,amssymb,amsfonts}

\usepackage{mathrsfs}
\usepackage{ulem}
\usepackage{amsopn} 

\usepackage{pdfpages}  

\renewcommand{\epsilon}{\varepsilon}
\renewcommand{\theta}{\vartheta}
\renewcommand{\kappa}{\varkappa}
\renewcommand{\rho}{\varrho} 
\renewcommand{\phi}{\varphi}

\newcommand{\comm}[1]{}

\newtheorem{theo}{Theorem}

\newcommand{\pd}[2]{\ifthenelse{\equal{\detokenize{#1}}{\detokenize{•}}}
						{\frac{\partial}{\partial #2}}
						{\frac{\partial #1}{\partial #2}}
					}
\newcommand{\Sum}[3]{\ifthenelse{\equal{\detokenize{#1}}{\detokenize{•}}}
							{\sum_{#2}{#3}}  
							{\sum^{#1}_{#2}{#3}}	
					}

\begin{document}
\title{\protect\vspace*{-2cm}Nonexistence of local conservation laws\\ for the generalized Swift-Hohenberg equation}
\author{Pavel Holba\\
Mathematical institute, Silesian University in Opava\\
Na Rybn\'\i{}\v{c}ku 1, 74601 Opava, Czech Republic}	
\maketitle	

\begin{abstract}
We prove that the generalized Swift--Hohenberg equation  with nonlinear right-hand side, a natural generalization of the Swift--Hohenberg equation arising in physics and describing inter alia pattern formation, has no nontrivial local conservation laws.
\end{abstract}



	\section*{Introduction}
Conservation laws are important in modern mathematical physics for many reasons \cite{i,kvv,olv93}. While the presence of an infinite series of conservation laws is usually a sign of integrability in the sense of soliton theory, cf.\ e.g.\ \cite{kvv, olv93, s18} and references therein, even existence of a finite number of conservation laws can be quite helpful in establishing the qualitative behavior of solutions, like e.g.\ preservation of the solution norm in a certain functional space, or of some important physical characteristics like energy or momentum, in the course of time evolution, cf.\ e.g.\ \cite{kvv, olv93}.
Notice that the search for conservation laws is a highly nontrivial task whose complexity grows significantly with the increase of number of independent variables and/or the order of the equation under study \cite{kvv, olv93}.\looseness=-1


Of course, an immediate consequence of the above is that it is also quite important to know that a certain equation has no (nontrivial) conservation laws at all, or, say, of order higher than a certain number, cf.\ e.g.\ \cite{i}, and below we prove just such a result, establishing nonexistence of nontrivial local conservation laws, for
the generalized Swift--Hohenberg equation in any number $n$ of space variables, that is,
%
	\begin{equation}
		u_t=A(\Delta)u+N(u),
		\label{gsh}
	\end{equation}
	where $A(\Delta)=\Sum{k}{i=0}{a_i\Delta^i}$, $k\geqslant 1$, $a_i$ are constants, $\Delta=\sum_{i=1}^n \partial^2/\partial x_i^2$ is the Laplace operator, and $N(u)$ is a smooth function of $u$. In what follows we make a blanket assumption that the polynomial $A(\Delta)$ is nonconstant so that (\ref{gsh}) is necessarily a PDE rather than an ODE for $u$.

The choice of name for equation (\ref{gsh}) is motivated by the fact that it is a natural generalization of the original Swift--Hohenberg equation \cite{sh}, which corresponds to the case when
\begin{equation}\label{she}
A(\Delta)=a (\Delta+b)^2+c,
\end{equation}
where $a,b,c$ are real constants, or, even more specifically, $a=b=1$, and has a number of important applications in physics. In particular, (\ref{gsh}) with $A$ given by (\ref{she}) serves as a model for the study of various issues in pattern formation, see e.g.\ \cite{f} and references therein.
 
Below we prove that \eqref{gsh} admits no local conservation laws if $N(u)$ satisfies $\partial^2 N/\partial u^2\neq 0$. Note that this is pretty much impossible to establish by direct computation, in particular  
because (\ref{gsh}) can, in view of freedom in choosing $A$, be of arbitrarily high even order.

\section{Preliminaries}
Following \cite{olv93}, we shall say that a {\sl differential function} is a smooth function of $x_1,\dots,x_n,t,u$ and finitely many $x$-derivatives of $u$.

Then a  {\sl local conserved vector} for (\ref{gsh}) is, cf.\ e.g.\ \cite{ps} and references therein, an $(n+1)$-tuple $(\rho,\sigma_1,\dots,\sigma_n)$ of differential functions that satisfies
\begin{equation}\label{cv}
D_t(\rho)+\sum\limits_{i=1}^n D_{x_i}\sigma_i=0 
\end{equation}
modulo (\ref{gsh}) and its differential consequences. 

We shall refer to the quantity $\delta\rho/\delta u$ as to the {\sl characteristic} of a conserved vector $(\rho,\sigma_1,\dots,\sigma_n)$. It is readily seen that for the case of (\ref{gsh}) this definition is equivalent to the standard one \cite{i,kvv,olv93}.

Here $D_t$ and $D_{x_i}$ are the so-called total derivatives and $\delta/\delta u$ is the variational derivative, see e.g.\ \cite{i,kvv,olv93,ps} for further details on those.

It is immediate that a linear combination of conserved vectors for (\ref{gsh}) is again a conserved vector for (\ref{gsh}), so conserved vectors for (\ref{gsh}) form a vector space.

A conserved vector $(\rho,\sigma_1,\dots,\sigma_n)$ for (\ref{gsh}) is said to be {\sl trivial} if its characteristic vanishes or, equivalently, if (\ref{cv}) holds for this conserved vector identically, without the need of invoking (\ref{gsh}) or its differential consequences, cf.\ e.g.\ \cite{i,kvv,olv93}.

Two conserved vectors for (\ref{gsh}) are said to be {\sl equivalent} if they differ by a trivial conserved vector, cf.\ \cite{kvv,ps}. 

A {\sl local conservation law} for (\ref{gsh}) is then defined, cf.\ e.g.\ \cite{kvv,ps}, as an equivalence class of conserved vectors with respect to the above equivalence relation.

It is readily seen, cf.\ e.g.\ \cite{i,olv93,ps}, that equivalent conserved vectors have the same characteristics, so the characteristic of a local conservation law for (\ref{gsh}), defined, cf.\ e.g.\ \cite{i,kvv,ps}, as a characteristic of any conserved vector from the respective equivalence class, is a well-defined quantity. Like for conserved vectors, a local conservation law is said to be {\sl trivial} if its characteristic identically vanishes. It can be shown that trivial conservation laws are pretty much of no interest for applications \cite{kvv,olv93}.


\section{Main result}
We are now in position to state our main result.
\begin{theo}\label{th6}
Equation \eqref{gsh} with $\partial^2 N/\partial u^2\neq 0$
has no nontrivial local conservation laws.
\end{theo}
	
\noindent{\sl Proof.} The necessary condition for a differential function, say $Q$, to be a characteristic of a local conservation law of \eqref{gsh} is readily seen, cf.\ e.g.\ \cite{i,kvv, olv93}, to take the form
	\begin{equation}
		D_t(Q)+\frac{\partial N}{\partial u}Q+\sum_{i=0}^{k} a_{i}{\tilde{\Delta}}^i (Q)=0,
		\label{char_def_eq}
	\end{equation}
where ${\tilde{\Delta}}=\sum_{i=1}^n D_{x_i}^2$.

It is easily verified that equation (\ref{gsh}), being an even-order evolution equation, belongs to a broader class of quasi-evolutionary equations that satisfy the conditions of Theorem 6 from \cite{i}. Moreover, the coefficients of (\ref{char_def_eq}) depend at most on $u$, and dependence on $u$ shows up only in zero-order term coefficient. Therefore, by the said theorem from \cite{i} for any local conservation law of (\ref{gsh}) its characteristic $Q$ depends at most on $t,x_1,\dots,x_n$ but not on $u$ and its derivatives.
		
With this in mind upon applying $\partial/\partial u$ to both sides of \eqref{char_def_eq} we get
	\begin{equation}
		\frac{\partial^2 N}{\partial u^2}Q=0,
	\end{equation}
which implies that if $\partial^2 N/\partial u^2\neq 0$ then $Q=0$, so (\ref{gsh}) can have only trivial local conservation laws, and the result follows. $\Box$

It is an interesting open problem to find out whether (\ref{gsh}) admits nontrivial differential coverings (see e.g.\ \cite{kvv} and references therein on those) and, if yes, whether (\ref{gsh}) would have nontrivial {\sl  nonlocal} conservation laws associated with these coverings.  




	
\subsection*{Acknowledgments}
This research was supported by the Specific Research grant SGS/6/2017 of the Silesian University in Opava.


\end{document}